\documentclass[copyright]{eptcs}
\usepackage{breakurl}        
\usepackage{amsmath,amssymb,amsthm}
\usepackage{prooftree}
\usepackage{display-figure}
\usepackage{uniqueness-macros}
\usepackage{stmaryrd}

\renewcommand{\subtype}{\ensuremath{\mathrel{<:}}}

\title{Uniqueness Typing for Resource Management in Message-Passing Concurrency}
\author{
  Edsko de Vries\thanks{This research was supported by SFI project SFI 06 IN.1 1898.}
  \institute{Trinity College Dublin, Ireland}
  \email{Edsko.de.Vries@cs.tcd.ie}
  \and
  Adrian Francalanza
  \institute{University of Malta}
  \email{Adrian.Francalanza@um.edu.mt}
  \and
  Matthew Hennessy\footnotemark[1]
  \institute{Trinity College Dublin, Ireland}
  \email{Matthew.Hennessy@cs.tcd.ie}
}

\def\titlerunning{Uniqueness Typing for Resource Management in Message Passing Concurrency}
\def\authorrunning{E. de Vries, A. Francalanza and M. Hennessy}

\begin{document}
\maketitle

\begin{abstract}
We view channels as the main form of resources in a message-passing programming
paradigm.  These channels need to be carefully managed in settings where
resources are scarce.  To study this problem, we extend the \pic with
primitives for channel allocation and deallocation and allow channels to be
reused to communicate values of different types.  Inevitably, the added
expressiveness increases the possibilities for runtime errors. We define a
substructural type system which combines uniqueness typing and affine typing to
reject these ill-behaved programs.
\end{abstract}

\section{Introduction}

Message-passing concurrency is a programming paradigm whereby shared memory is
prohibited and process interaction is limited to explicit message
communication. This concurrency paradigm forms the basis for a number of
process calculi such as the \pic \cite{Milner:Pi} and has been adopted by
programming languages such as the actor based language Erlang
\cite{Armstrong:Erlang}.   

Message-passing concurrency often abstracts away from resource management and
programs written at this abstraction level exhibit poor resource awareness. In
this paper we study ways of improving this shortcoming. Specifically, we
develop a statically typed extension of the \pic in which resources, i.e.
channels, can be reused at varying types and unwanted resources can be safely
deallocated. 

Idiomatic \pic processes are often characterized by wasteful
use-once-throw-away channels \cite{Milner:Pi,KobayashiPT:linearity}.  Consider
the following two \pic process definitions
\begin{align*}
\ptit{timeSrv} &\deftri \recX{\piin{\textit{getTime}}{x}{\pioutB{x}{\textit{hr}, \textit{min}}{X}}} \\ 
\ptit{dateSrv} &\deftri \recX{\piin{\textit{getDate}}{x}{\pioutB{x}{\textit{year}, \textit{mon}, \textit{day}}{X}}}
\end{align*} 
\ptit{timeSrv} defines a server that repeatedly waits on a channel named
\textit{getTime} to dynamically receive a channel name, represented by
the bound variable $x$, and then replies with the current time on $x$.
\ptit{dateSrv} is a similar service which returns the current date.  An
idiomatic \pic client definition is
\begin{equation*}
  \ptit{client}_0 \deftri \rest{\textit{ret}_1}{\,\pioutB{\textit{getTime}}{\textit{ret}_1}{\piinB{\textit{ret}_1}{y_\textit{hr}, y_\textit{min}}{\,\rest{\textit{ret}_2}{\,\pioutB{\textit{getDate}}{\textit{ret}_2}{\piinB{\textit{ret}_2}{z_\mathit{year}, z_\mathit{mon}, z_\mathit{day}}{\,P}}}}}}
\end{equation*}
$\ptit{client}_0$ uses two distinct channels $\textit{ret}_1$ and
$\textit{ret}_2$ as return channels to query the time and date servers, and
then continues as process $P$ with the values obtained. These return
channels are scoped (private), to preclude interference from other clients
concurrently querying the servers.  

From a resource management perspective it makes pragmatic sense to try and
reduce the number of channels used and use \emph{one} channel to communicate
with both the time server and the date server. 
\begin{equation*} \ptit{client}_1 \deftri
\rest{\textit{ret}}{\,\pioutB{\textit{getTime}}{\textit{ret}}{\piinB{\textit{ret}}{y_\mathit{hr}, y_\mathit{min}}{\,\pioutB{\textit{getDate}}{\textit{ret}}{\piinB{\textit{ret}}{z_\mathit{year}, z_\mathit{mon}, z_\mathit{day}}{\,P}}}}}
\end{equation*}
From a typing perspective, this reuse of the same channel entails \emph{strong
update} on the channel: that is, reuse of a channel to communicate values of
different types. Strong update must be carefully controlled; for instance, an
attempt to use one channel to communicate with both servers in parallel is
unsafe:
\begin{equation*}
  \ptit{client}_\text{err} \deftri \restB{\textit{ret}}{\,\pioutB{\textit{getTime}}{\textit{ret}}{\piinB{\textit{ret}}{y_\textit{hr}, y_\textit{min}}{\,P_1}} \paralS {\,\pioutB{\textit{getDate}}{\textit{ret}}{\piinB{\textit{ret}}{z_\mathit{year}, z_\mathit{mon}, z_\mathit{day}}{\,P_2}}}}
\end{equation*}
Standard \pic type systems accept only $\ptit{client}_0$ and rule out both
$\ptit{client}_1$ and $\ptit{client}_\text{err}$. However, $\ptit{client}_1$ is
safe because the communication with the date server happens \emph{strictly after}  the
communication with the time server, and also because the time server will only use the
return channel \emph{once}. 

Adequate resource management also requires precise descriptions of when
resources are allocated and existing ones are disposed.  The characteristic
scoping construct $\rest{c}{P}$ is  unfit for this purpose and should be used
only as a bookkeeping construct delineating name scoping (which evolves during
computation through scope extrusion) . One reason against such an
interpretation  is the structural rule 
\begin{align*}
\rtit{(sScp)} &\qquad P \steq  \rest{c}{P}  \quad \text{whenever}\quad c\not\in\fn{P}
\end{align*}
whose symmetric nature would entail implicit garbage collection of channels,
but also the possibility of unfettered spurious channel allocations. Another
reason against this interpretation is the fact that the \pic semantics does not
specify whether, in a process such as $\ptit{client}_0$, channel
$\textit{ret}_2$ is allocated before or after the input on $\textit{ret}_1$.
This problem becomes more acute when scoping occurs inside recursive
definitions.  We address this shortcoming by introducing an explicit allocation
construct $\alloc{x}{P}$.  When the allocation is executed,  a new channel $c$
is created at runtime and the $\alloc{x}{P}$ changes to
$\rest{c}{P\subC{c}{x}}$.  Dually, we also extend the calculus with an
explicit \emph{deallocation} operator $\free{c}{P}$. We can then rewrite the
client as: 
\begin{equation*}
  \ptit{client}_2 \deftri \alloc{x}{\pioutB{\textit{getTime}}{x}{\piinB{x}{y_\mathit{hr}, y_\mathit{min}}{\pioutB{\textit{getDate}}{x}{\piinB{x}{z_\mathit{year}, z_\mathit{mon}, z_\mathit{day}}{\free{x}{P}}}}}}
\end{equation*}
Inevitably, the added expressiveness of this extended \pic increases the
possibilities for runtime errors like value mismatch during communication and
usage of channels which have been deallocated.  

We define a type system which rejects processes that are unsafe; the type
system combines uniqueness typing \cite{barendsen:functional} and affine typing
\cite{KobayashiPT:linearity}, while permitting value coercion across these
modes through subtyping. Uniqueness typing gives us \emph{global guarantees},
simplifying local reasoning when typing both strong updates and safe
deallocations. Uniqueness typing can be seen as dual to affine typing
\cite{harrington:uniquenesslogic}, and we make essential use of this duality to
allow uniqueness to be temporarily violated.

\section{The \picR}

\figref{fig:syntax-ext} shows the syntax for the \picr. The language is the
standard \pic extended with primitives for explicit channel allocation and
deallocation; moreover, channel scoping records whether a channel is allocated
($\top$) or deallocated ($\bot$).  The syntax assumes two separate denumerable
sets of channel names $c,\,d\in \Names$ and variables $x,\,y\in\Vars$, and lets
identifiers $u,\,v \in \Names\cup\Vars$ range over both.  The input and channel
allocation constructs are binders for variables $\vec{x}$ and $x$ \resp,
whereas scoping is a binder for names (\ie $c$).  The syntax also assumes a
denumerable set of process variables $X,Y \in \PVars$ which are bound by the
recursion construct. 

\begin{display}{Polyadic \picr syntax}{fig:syntax-ext}
\begin{equation*}
\begin{array}{l@{\hspace{1ex}}r@{\hspace{1ex}}lllllllllll}
P, Q & \bnfdef & \piout{u}{\vec{v}}{P} & \textsl{(output)}    & \bnfsep & \piin{u}{\vec{x}}{P} & \textsl{(input)}    \\ 
     & \bnfsep & \inert                & \textsl{(nil)}       & \bnfsep & \match{u}{v}{P}{Q}   & \textsl{(match)}    \\       
     & \bnfsep & \recX{P}              & \textsl{(recursion)} & \bnfsep & X                    & \textsl{(process variable)} \\              
     & \bnfsep & P \paral Q            & \textsl{(parallel)}  & \bnfsep & \restT{c}{s}{P}      & \textsl{(stateful scoping)} \\
     & \bnfsep & \alloc{x}{P}          & \textsl{(allocate)}  & \bnfsep & \free{u}{P}          & \textsl{(deallocate)}       \\
\end{array}
\end{equation*}
\end{display}

Channels are stateful (allocated, \salloc, or deallocated, \sdalloc) and
process semantics is defined over configurations, $\langle \st, P\rangle$ where
$\st\in\Sigma  : \Chans \pmap \sset{\salloc, \sdalloc}$ describes the state of
the free channels in $P$, and stateful scoping \restT{c}{s}{P} describes the
state of scoped channels.  A tuple $\langle \st, P\rangle$ is a configuration
whenever $\fn{P}\subseteq\dom{\st}$ and is denoted  as \confSP. We say that a
configuration \confSP\ is closed whenever $P$ does not have any free variables.
\figref{fig:pir-reduction} defines contexts over configuration where
$\ctxtGen{\confSP}$ denotes the application of a context $\ctxt$ to a
configuration \confSP.  In the case where a context scopes a name $c$ the
definition extracts the state relating to $c$ from \st{} and associates it with
$c$. For example,
\begin{equation*}
\ctxtR{c}{\ctxtEmp{\;\conf{\envmap{c}{\top},\envmap{d}{\top}}{\pioutA{c}{d}}\;}}
=
\conf{\envmap{d}{\top}}{\restT{c}{\top}{\pioutA{c}{d}}}
\end{equation*}

The reduction relation is defined as the least contextual relation over closed
configurations satisfying the rules in \figref{fig:pir-reduction}, using a
standard \pic structural equivalence relation $(\equiv)$.  Communication
(\rtit{rCom}) requires the communicating channel to be allocated. Allocation
(\rtit{rAll}) creates a private allocated channel and substitutes it for the
bound variable of the allocation construct in the continuation; the condition
$c\not\in\dom{\st}$ ensures that $c$ is fresh in $P$. Deallocation (\rtit{rFree})
is the only construct that changes the visible state of a configuration, \st. 

\begin{display}{Contexts, Reduction Rules and Error Predicates}{fig:pir-reduction}
\textbf{Contexts}
\begin{equation*}
  \ctxt \bnfdef \quad \ctxtEmp{-} \quad | \quad \ctxtPar{\ctxt}{P} \quad | \quad \ctxtPar{P}{\ctxt} \quad | \quad \ctxtR{c}{\ctxt}\\[5pt]
\end{equation*}
\begin{equation*}
\begin{array}{rll}
  \ctxtEmp{\confSP} & \deftxt \confSP \\
  \ctxtPar{\ctxtGenN{\context}{\confSP}}{Q} & \deftxt \conf{\st'}{(P'\paral Q)} & \text{if } \ctxtGenN{\context}{\confSP}=\conf{\st'}{P'}\\
  \ctxtPar{Q}{\ctxtGenN{\context}{\confSP}} & \deftxt \conf{\st'}{(Q\paral P')} & \text{if } \ctxtGenN{\context}{\confSP}=\conf{\st'}{P'}\\
  \ctxtR{c}{\ctxtGenN{\context}{\confSP}} & \deftxt \conf{\st'}{\restT{c}{s}{P'}} & \text{if } \ctxtGenN{\context}{\confSP}=\conf{\st',\envmap{c}{s}}{P'}
\end{array}
\end{equation*}

\textbf{Reduction Rules}\\

\begin{tabular}{ll}
\begin{prooftree}
\st(c) = \salloc
\justifiedBy{\rtit{rCom}}
\confS{\,\pioutA{c}{\vec{b}} \paral \piin{c}{\vec{x}}{P}}\reduc \confS{\,P\subC{\vec{b}}{\vec{x}}}
\end{prooftree}
&
\begin{prooftree}
\strut
\justifiedBy{\rtit{rRec}}
\confS{\recX{P}}\reduc \confS{P\subC{\recX{P}}{X}}
\end{prooftree} 
\\[2em]
\begin{prooftree}
\strut
\justifiedBy{\rtit{rThen}}
\confS{\match{c}{c}{P}{Q}}\reduc \confS{P}
\end{prooftree}
&
\begin{prooftree}
c \neq d
\justifiedBy{\rtit{rElse}}
\confS{\match{c}{d}{P}{Q}}\reduc \confS{Q}
\end{prooftree}
\\[2em]
  \begin{prooftree}
    c\not\in\dom{\st}
    \justifiedBy{\rtit{rAll}}
    \confS{\alloc{x}{P}}\reduc \confS{\restTB{c}{\salloc}{P\subC{c}{x}}}
  \end{prooftree}
&
  \begin{prooftree}
    \strut
    \justifiedBy{\rtit{rFree}}
    \conf{\st,c\!:\!\salloc}{\,\free{c}{P}}\reduc \conf{\st,c\!:\!\sdalloc}{P} 
  \end{prooftree}
\\[2em]
\multicolumn{2}{c}{
\begin{prooftree}
    P \steq P' \qquad \confS{P'} \reduc  \confS{Q'} \qquad Q'\steq Q
    \justifiedBy{\rtit{rStr}}
    \confSP\reduc \confS{Q} 
  \end{prooftree}
}\\[2em]
\end{tabular}

\textbf{Error Reduction Rules}

\begin{equation*}
\begin{prooftree}
|\vec{d}| \neq |\vec{x}|
\justifiedBy{eAty}
\confS{\piout{c}{\vec{d}}{P} \paralS \piin{c}{\vec{x}}{Q} \reducl{\text{err}}}
\end{prooftree}
\quad
\begin{prooftree}
\st(c)=\sdalloc
\justifiedBy{eOut}
\confS{\piout{c}{\vec{d}}{P}} \reducl{\text{err}}
\end{prooftree}
\quad
\begin{prooftree}
\st(c)=\sdalloc
\justifiedBy{eIn}
\confS{\piin{c}{\vec{x}}{Q}} \reducl{\text{err}}
\end{prooftree}
\quad 
\begin{prooftree}
P\steq Q \quad\confS{Q} \reducl{\text{err}}
\justifiedBy{eStr}
\confSP \reducl{\text{err}}
\end{prooftree}
\end{equation*}
\end{display}

\figref{fig:pir-reduction} also defines error reductions as the least
contextual relation satisfying the rules for $\reducl{\text{err}}$.   These
rules capture errors resulting from arity mismatch and attempts to communicate
on deallocated channels.  In particular arity mismatch can come from
unconstrained use of strong updates such as in the case of
$\ptit{client}_\text{err}$ in the Introduction.

\section{Type System}
\label{sec:type-system}

\subsection{The typing relation}

The type language is defined in Figure~\ref{fig:types} and the typing rules are
given in Figure~\ref{fig:typingrules}. The typing relation over processes takes
the usual shape: $\tprocEP$ read as ``$P$ is well-typed under the typing
assumptions in $\env$''. Typing environments are multisets of pairs of
identifiers and types; we do not \emph{a priori} make the assumption that they are
partial functions (Section~\ref{sec:Consistency}). This relation is extended to
configurations as $$\tcnf{\env}{\confSP}$$ by requiring that the process is
well-typed in $\env$, all channels in $\env$ are allocated in $\sigma$, and
$\env$ is a partial function (\ie for every identifier in $P$ there is exactly one
typing assumption in $\env$). 

\subsection{The type language}

\begin{display}{Type language}{fig:types}
\begin{equation*}
\begin{array}{lrll}
\tV & \bnfdef & \chantyp{\lst{\tV}}{\aV} & \text{(channel type)} \\
    & \bnfsep & \proctyp                 & \text{(process)} 
\end{array}\qquad 
\begin{array}{lrll}
\aV & \bnfdef & \affine       & \text{(affine)} \\
    & \bnfsep & \unrestricted & \text{(unrestricted)} \\
    & \bnfsep & \unique{i}    & \text{(unique after $i$ steps, $i \in \mathbb{N}$)}  
\end{array}
\end{equation*}
\end{display}

The core type of our system is the channel type, $\chantyp{\lst{\tV}}{a}$,
consisting of an $n$-ary tuple of types $\lst{\tV}$ describing the values carried
over the channel and an \emph{attribute} $a$ which gives usage information about the
channel (a channel is \textit{used} when communication takes place across the
channel). This attribute can take one of three forms:
\begin{itemize}
\item A channel of type $\chantyp{\lst{\tV}}{\unrestricted}$ is an
\textit{unrestricted} channel; such type assumptions correspond to type
assumptions of the form $[\lst{\tV}]$ in non-substructural type systems.
\item A channel of type $\chantyp{\lst{\tV}}{\affine}$ is \textit{affine}, and
comes with an obligation: it can be used at most once.
\item A channel of type $\chantyp{\lst{\tV}}{\unique{i}}$ comes with a
guarantee that it is \textit{unique} after $i$ actions; we
abbreviate the type  $\chantyp{\lst{\tV}}{\unique{0}}$ of channels that are
unique \textit{now} to $\chantyp{\lst{\tV}}{\uniqueNow}$. 
Unique channels can be used to describe instances where only one process has
access to  (owns) a channel. Accordingly, strong update and deallocation is
safe for unique channels.
\end{itemize}

\subsection{Structural rules}

\begin{display}{Typing rules}{fig:typingrules}
\textbf{Logical rules}\\

\begin{tabular}{llll}
\quad
&
\begin{prooftree}
\tprocP{\env, \envmap{u}{\chantyp{\tVlst}{\aV-1}}, \lst{\envmap{x}{\tV}}} 
\justifiedBy{tIn}
\tproc{\env, \envmap{u}{\chantypA{\tVlst}}}{\piin{u}{\vec{x}}{P}}  
\end{prooftree}
&
\begin{prooftree}
\tprocP{\env, \envmap{u}{\chantyp{\tVlst}{\aV-1}}}  
\justifiedBy{tOut}
\tproc{\env, \envmap{u}{\chantyp{\tVlst}{\aV}}, \lst{\envmap{v}{\tV}}}{\piout{u}{\vec{v}}{P}}
\end{prooftree}
 &
\begin{prooftree}
\tprocP{\env_1} \qquad
\tproc{\env_2}{Q} 
\justifiedBy{tPar}
\tproc{\env_1, \env_2\,}{\,P \paral Q}
\end{prooftree}
\\[2em]
&
\begin{prooftree}
u, v \in \Gamma \qquad
\tprocE{P} \qquad
\tprocE{Q}
\justifiedBy{tIf}
\tprocE{\match{u}{v}{P}{Q}}
\end{prooftree} 
 &
\begin{prooftree}
\tprocP{\env^\unrestricted, \envmap{X}{\proctyp}} 
\justifiedBy{tRec}
\tproc{\env^\unrestricted}{\recX{P}} 
\end{prooftree} 
 &
\begin{prooftree}
\strut
\justifiedBy{tVar}
\tproc{\envmap{X}{\proctyp}}{X}
\end{prooftree}
\\[2em]
&
\begin{prooftree}
\tprocP{\env,\envmap{x}{\chantyp{\lst{\tV}}{\uniqueNow}}}
\justifiedBy{tAll}
\tprocE{\alloc{x}{P}}
\end{prooftree}
 &
\begin{prooftree}
\tprocEP
\justifiedBy{tFree}
\tproc{\env,\envmap{u}{\chantyp{\lst{\tV}}{\uniqueNow}}}{\free{u}{P}}
\end{prooftree}
&
\begin{prooftree}
\strut
\justifiedBy{tNil}
\tproc{\emptyset}{\inert}
\end{prooftree}   
\\[2em]
&
\begin{prooftree}
\tprocP{\env,\envmap{c}{\tV}}
\justifiedBy{tRst1}
\tprocE{\restT{c}{\salloc}{P}}
\end{prooftree} 
 &
\begin{prooftree}
\tprocEP
\justifiedBy{tRst2}
\tprocE{\restT{c}{\sdalloc}{P}}
\end{prooftree} 
 \\[2em]
\end{tabular}

where $\env^\unrestricted$ is an environment containing only unrestricted
assumptions and all bound variables are fresh.\\[1em]
\textbf{Structural rules}
\begin{equation*}
\begin{prooftree}
\tV = \tV_1 \circ \tV_2 \quad
\tprocP{\env, \envmap{u}{\tV_1}, \envmap{u}{\tV_2}} 
\justifiedBy{tCon}
\tprocP{\env, \envmap{u}{\tV}}
\end{prooftree} \qquad
\begin{prooftree}
\tprocP{\env}
\justifiedBy{tWeak}
\tprocP{\env, \envmap{u}{\tV}}
\end{prooftree} 
\end{equation*}
\begin{equation*}
\begin{prooftree}
\tprocP{\env,\envmap{u}{\tV_2}} \quad
\tV_1 \subtype \tV_2
\justifiedBy{tSub}
\tprocP{\env,\envmap{u}{\tV_1}}
\end{prooftree} \qquad
\begin{prooftree}
\tprocP{\env,\envmap{u}{\chantyp{\lst{\tV_2}}{\uniqueNow}}} 
\justifiedBy{tRev}
\tprocP{\env,\envmap{u}{\chantyp{\lst{\tV_1}}{\uniqueNow}}}
\end{prooftree}
\end{equation*}

\textbf{Typing configurations}\\[7pt]
\begin{equation*}
\begin{prooftree}
\forall c \in\dom{\env}.\, \st(c) = \salloc \qquad 
\tprocEP \qquad
\Gamma \text{ is a partial map}
\justifiedBy{tConf}
\tcnf{\env}{\confSP}
\end{prooftree}  
\end{equation*}  

\textbf{Channel usage}
\begin{align*}
  \env, \envmap{c}{\chantyp{\lst{\tV}}{\aV-1}} & \deftxt
  \begin{cases}
    \env & \text{if }\,\aV=\affine\\
    \env, \envmap{c}{\chantyp{\vec{\tV}}{\unrestricted}} & \text{if }\,\aV=\unrestricted\\
    \env, \envmap{c}{\chantyp{\vec{\tV}}{\unique{i}}} &  \text{if }\,\aV=\unique{i + 1}
  \end{cases}
\end{align*}

\textbf{Type splitting}
\begin{equation*}
\begin{prooftree}
\justifiedBy{pUnr}
\chantyp{\tVlst}{\unrestricted} = \chantyp{\tVlst}{\unrestricted} \circ \chantyp{\tVlst}{\unrestricted}
\end{prooftree} \qquad
\begin{prooftree}
\justifiedBy{pProc}
\proctyp = \proctyp \circ \proctyp
\end{prooftree} \qquad 
\begin{prooftree}
\justifiedBy{pUnq}
\chantyp{\tVlst}{\unique{i}} = \chantyp{\tVlst}{\affine} \circ \chantyp{\tVlst}{\unique{i+1}}
\end{prooftree}
\end{equation*}
\textbf{Subtyping}
\begin{equation*}
\begin{prooftree}
\phantom{\aV_1 \subtype \aV_2}
\justifiedBy{sIndx}
\unique{i} \subtype \unique{i+1}
\end{prooftree} \qquad 
\begin{prooftree}
\phantom{\aV_1 \subtype \aV_2}
\justifiedBy{sUnq}
\unique{i+1} \subtype \unrestricted
\end{prooftree} \qquad 
\begin{prooftree}
\phantom{\aV_1 \subtype \aV_2}
\justifiedBy{sAff}
\unrestricted \subtype \affine 
\end{prooftree} \qquad 
\begin{prooftree}
\aV_1 \subtype \aV_2
\justifiedBy{sTyp}
\chantyp{\tVlst}{\aV_1} \subtype \chantyp{\tVlst}{\aV_2}
\end{prooftree} 
\end{equation*}
\end{display}

Since the type system is substructural, usage of type assumptions must be
carefully controlled, and the logical rules do not allow to use an assumption
more than once. Operations on the typing environment are described separately
by the structural rules of \figref{fig:typingrules}.  Although the subtyping
relation is novel because it combines uniqueness subtyping (\rtit{sUnq}) with
affine subtyping (\rtit{sAff}) making them \emph{dual} with respect to
unrestricted types, the subtyping and weakening structural rules are standard.  

Rule \textsc{tCon}  contracts an assumption of the form $\envmap{u}{\tV}$, as
long as $\tV$ can be \textit{split} as $\tV_1$ and $\tV_2$. Unrestricted
assumptions can be split arbitrarily (\rtit{pUnr} and \rtit{pProc}); affine
assumptions cannot be split at all. An assumption
$\envmap{u}{\chantyp{\vec{\tV}}{\uniqueNow}}$ about a unique channel can be
split as an affine assumption $\envmap{u}{\chantyp{\vec{\tV}}{\affine}}$ and an
assumption about a channel that is unique after one action
$\envmap{u}{\chantyp{\vec{\tV}}{\unique{1}}}$---an action using the latter
assumption must be coupled with a co-action on the affine channel, and since
the affine assumption can only be used once it is sound to assume that the
channel is unique again after the action has happened. More generally, a
channel which is unique after $i$ actions can be split into an affine
assumption and a channel which is unique after $(i+1)$ actions, and splitting
is defined in such a way that the number of affine assumptions for a channel
never exceeds the index $i$ of the corresponding unique assumption. 

In particular when the index is 0 no other assumptions about that channel can
exist in the typing environment. This means that if a process can be typed
using a unique assumption for a channel, no other process has access to that
channel.  The last structural rule, \textsc{tRev}, makes use of this fact to
allow strong updates (``revision'') to channels as long as they are unique.   

\subsection{Logical rules}

The rules for input and output (\rtit{tOut} and \rtit{tIn}) decrement the
attribute of the channel, \ie they \emph{count} usage.  This operation is
denoted by $\env, \envmap{c}{\chantyp{\vec{\tV}}{a-1}}$ in
\figref{fig:typingrules} and states that affine assumptions can only be used
once, unrestricted assumptions can be used an arbitrary number of types, and if
a channel is unique after $i\!+\!1$ actions, then it will be unique after $i$
actions once the action has been performed.

Moreover, \rtit{tOut} requires separate typing assumptions for each of the
channels that are sent. The attributes on these channels are not decremented,
because no action has been performed on them; instead, the corresponding
assumptions are handed over to the parallel process receiving the message. If the sending process
wants to use any of these channels in its continuation ($P$) it must split the
corresponding assumptions first. 

Recursive processes must be typed in an environment that contains only
unrestricted channels (\rtit{tRec}). This is reasonable since recursion can be
used to define processes with an unbounded number of parallel uses of some
channel. Nevertheless, it is not as serious a restriction as it may seem ,as
recursive processes can still send, receive and allocate unique channels. For
instance, the following process models an ``infinite heap'' that keeps
allocating new channels and sends them across a channel
$\envmap{\textit{heap}}{\chantyp{\chantyp{\tV}{\uniqueNow}}{\unrestricted}}$:
\begin{equation*}
\ptit{infHeap} \deftri \recX{\alloc{x}{\piout{\mathit{heap}}{x}{X}}}
\end{equation*}
As expected, allocation introduces unique channels (\rtit{tAlloc}) and only
unique channels can be deallocated (\rtit{tFree}). Finally, a typing assumption
is introduced in the typing environment for locally scoped names only if the
corresponding channel is allocated (\rtit{tRst1} and \rtit{rRst2}). 

\subsection{Consistency}
\label{sec:Consistency}

When we take a bird's eye-view of a system, every channel has exactly one type.
In the definition of $\tcnf{\env}{\confSP}$, we therefore restrict the
environment to have at most one assumption about every channel: we require that
$\env$ is a partial function. Consequently, we  state type safety and
subject reduction lemmas with respect to environments that are partial
functions.

Nevertheless, when two processes both need a typing assumption relating to the
same channel $c$, there need to be two separate assumptions in the typing
environment (\cf \rtit{tPar} in \figref{fig:typingrules}). These two
assumptions need not be identical;  for example, an assumption $c :
\chantyp{}{\uniqueNow}$ can be split as two assumptions $$c :
\chantyp{}{\unique{1}}, c : \chantyp{}{\affine}$$ 

A \emph{consistent} environment is defined to be an environment that can be
obtained by applying any of the structural rules (contraction, weakening,
subtyping or revision) from an environment which is a partial function. It
follows that any process that can be typed in a consistent environment can also
be typed in a environment that is a partial function. The reader may therefore
wonder why we do not restrict the typing relation to partial functions.  It
turns out that even if a process can be typed in a consistent environment, some
of its subprocesses might have to be typed in an inconsistent environment.  As
an example, consider the typing derivation
\begin{equation*}
\begin{prooftree}
\[
  \[
    \justifiedBy{tOut}
    \envmap{a}{\chantyp{\chantyp{}{\affine}}{\unrestricted}}, \envmap{u}{\chantyp{}{\affine}} \vdash \pioutA{a}{u} 
  \]
  \[
    \[
      \envmap{a}{\chantyp{\chantyp{}{\affine}}{\unrestricted}}, \envmap{u}{\chantyp{}{\uniqueNow}}, x : \chantyp{}{\affine} \vdash \freeA{u} \paral \pioutA{a}{x} 
      \justifiedBy{tIn}
      \envmap{a}{\chantyp{\chantyp{}{\affine}}{\unrestricted}}, \envmap{u}{\chantyp{}{\unique{1}}}, x : \chantyp{}{\affine} \vdash \piinBB{u}{()}{\freeA{u} \paral \pioutA{a}{x}}
    \]
    \justifiedBy{tIn}
    \envmap{a}{\chantyp{\chantyp{}{\affine}}{\unrestricted}}, \envmap{u}{\chantyp{}{\unique{1}}} \vdash \piin{a}{x}{\piinBB{u}{()}{\freeA{u} \paral \pioutA{a}{x}}}
  \]
  \justifiedBy{tPar}
  \envmap{a}{\chantyp{\chantyp{}{\affine}}{\unrestricted}}, \envmap{a}{\chantyp{\chantyp{}{\affine}}{\unrestricted}}, \envmap{u}{\chantyp{}{\affine}}, \envmap{u}{\chantyp{}{\unique{1}}} \vdash \pioutA{a}{u} \paral \piin{a}{x}{\piinBB{u}{()}{\freeA{u} \paral \pioutA{a}{x}}}
\]
\justifiedBy{tCon \textnormal{(twice)}}
\envmap{a}{\chantyp{\chantyp{}{\affine}}{\unrestricted}}, \envmap{u}{\chantyp{}{\uniqueNow}} \vdash \pioutA{a}{u} \paral \piin{a}{x}{\piinBB{u}{()}{\freeA{u} \paral \pioutA{a}{x}}}
\end{prooftree}
\end{equation*}
This is a valid typing derivation, and moreover the typing environment used at
every step is consistent. But now consider what happens after this process
takes a reduction step:
\begin{equation*}
\begin{prooftree}
\[
  \envmap{u}{\chantyp{}{\uniqueNow}}, u : \chantyp{}{\affine} \vdash \freeA{u} \paral \pioutA{a}{u} 
  \justifiedBy{tIn}
  \envmap{u}{\chantyp{}{\unique{1}}}, u : \chantyp{}{\affine} \vdash \piinBB{u}{()}{\freeA{u} \paral \pioutA{a}{u}}
\]
\justifiedBy{tCon}
\envmap{u}{\chantyp{}{\uniqueNow}} \vdash \piinBB{u}{()}{\freeA{u} \paral \pioutA{a}{u}}
\end{prooftree}
\end{equation*}
The tail of this process looks suspicious as it attempts to free $u$ while
simultaneously sending it on $a$. Indeed, $\freeA{u} \paral \pioutA{a}{u}$ can
only be typed in an inconsistent environment
$\envmap{u}{\chantyp{}{\uniqueNow}}, u : \chantyp{}{\affine}$.  Nevertheless,
the fact that this process is typeable is not a violation of type safety. The
assumption $\envmap{u}{\chantyp{}{\uniqueNow}}$ tells us that there are no
processes that output on $u$ so that the input on $u$ is blocked: the tail of
the process will never execute.

Thus, when an environment $$\env, \envmap{c}{\chantyp{\vec{\tV}}{a}},
\envmap{c}{\chantyp{\vec{\tV}}{a'}}$$ (e.g., $\envmap{u}{\chantyp{}{\unique{1}}},
u : \chantyp{}{\affine}$) is consistent, it may be the case that $$\env,
\envmap{c}{\chantyp{\vec{\tV}}{a-1}}, \envmap{c}{\chantyp{\vec{\tV}}{a'}}$$ (e.g.,
$\envmap{u}{\chantyp{}{\uniqueNow}}, u : \chantyp{}{\affine}$) is \emph{inconsistent}:
this means that the tails of input or output processes may have to be typed
under inconsistent environments, even when the larger process is typed in a
consistent environment.

However, communication in the \pic provides synchronization points: when a
communication happens, two processes will \textit{both} start executing their
tail processes. The following lemma says that if this happens, the resulting
overall system (of both processes) is still typeable in a consistent
environment; in fact, both must be derivable from the \textit{same}
partial function. This lemma is crucial in the subject reduction proof.

\begin{lemma}
Let $\env, \envmap{u}{\chantyp{\vec{\tV}}{a_1}},
\envmap{u}{\chantyp{\vec{\tV}}{a_2}}$ be a consistent environment, i.e.,
derivable from a partial function $\env'$ by applying structural rules. Then
$\env, \envmap{u}{\chantyp{\vec{\tV}}{a_1-1}},
\envmap{u}{\chantyp{\vec{\tV}}{a_2-1}}$ is derivable from the same environment
$\env'$, and is therefore consistent. 
\end{lemma}

\subsection{Soundness}

We prove soundness of the type system in the usual way.

\begin{theorem}[Type safety]
If $\tcnf{\env}{\confSP}$ then $P \nrightarrow^\text{err}$.
\label{thm:safety}
\end{theorem}

\begin{theorem}[Subject reduction]
If \tcnf{\env}{\confSP} and $\confSP \rightarrow
\conf{\st'}{P'}$ then there exists a environment $\env'$ such that 
$\tcnf{\env'}{\conf{\st'}{P'}}$.   
\end{theorem}

\noindent
Taken together these two theorems imply that a well-typed process will not have
any runtime errors. The proofs of these theorems can be found in the
accompanying technical appendix \cite{devries:linearity-appendix}.

\subsection{Examples}

The systems $\ptit{client}_i \paral \ptit{timeSrv} \paral \ptit{dateSrv}$ for
$i\in \sset{0,1,2}$ can all be typed in our type system, whereas
$\ptit{client}_\text{err}$ is rejected because type splitting enforces a common
object type (\cf \rtit{pUnr}, \rtit{pUnq} in \figref{fig:typingrules}.) For
convenience, we here recall $\ptit{client}_2$  and consider how it is typed,
assuming $x$ is not free in $P$: 
\begin{equation*}
  \ptit{client}_2 \deftri \alloc{x}{\pioutB{\textit{getTime}}{x}{\piinB{x}{y_\mathit{hr}, y_\mathit{min}}{\pioutB{\textit{getDate}}{x}{\piinB{x}{z_\mathit{year}, z_\mathit{mon}, z_\mathit{day}}{\free{x}{P}}}}}}
\end{equation*}
Assuming an environment with
\envmap{\textit{getTime}}{\chantyp{\chantyp{\tV_1,\tV_2}{\affine}}{\unrestricted}}
and
\envmap{\textit{getDate}}{\chantyp{\chantyp{\tV_3,\tV_4,\tV_5}{\affine}}{\unrestricted}},
$\ptit{client}_2$ types as follows.  Rule \rtit{tAll} assigns the unique type
\chantyp{\tV_1, \tV_2}{\uniqueNow} to variable $x$ and the
structural rule \rtit{tCon} then splits this unique assumption in two using
\rtit{pUnq}. Rule \rtit{tOut} uses the affine assumption for $x$ for the output
argument and the unique-after-one assumption to type the continuation.  Rule
\rtit{tIn} restores the uniqueness of $x$ for the continuation of the input
after decrementing the uniqueness index, at which point \rtit{tRev} is applied
to change the object type of $x$ from pairs of integers (for time) to triples
of integers (for dates). The pattern of applying \rtit{tCon}, \rtit{tOut} and
\rtit{tIn} repeats, at which point $x$ is unique again and can be safely
deallocated by \rtit{tFree}. 

Uniqueness allows us to typecheck a third client variation manifesting (explicit)
ownership transfer. Rather than allocating a new channel, $\ptit{client}_3$
requests a channel from a heap of channels and returns the channel to the heap
when it no longer needs it, thereby reusing channels \emph{across} clients.
\begin{equation*}
\ptit{client}_3 \deftri \piin{\textit{heap}}{x}{\pioutB{\textit{getTime}}{x}{\piinB{x}{y_\mathit{hr}, y_\mathit{min}}{\pioutB{\textit{getDate}}{x}{\piinB{x}{z_\mathit{year}, z_\mathit{mon}, z_\mathit{day}}{\piout{\textit{heap}}{x}{P}}}}}}
\end{equation*}
With the assumption
$\envmap{\textit{heap}}{\chantyp{\chantyp{\tV}{\uniqueNow}}{\unrestricted}}$
typing the system below is analogous to the previous client typings.
\begin{align}\label{eq:2}
   &\ptit{client}_3 \paral \ptit{client}_3 \paral \ptit{client}_3 \paral \ptit{timeSrv} \paral \ptit{dateSrv} \paral \rest{c}{\pioutAB{\textit{heap}}{c}} 
\end{align}

In $\ptit{client}_2$ and $\ptit{client}_3$, substituting $\piout{x}{()}{Q}$ for
$P$  makes the clients unsafe (they perform \resp  deallocated-channel usage
and mismatching communication).  Both clients would be rejected by the type
system because the use of  $x$ in $\piout{x}{()}{Q}$ requires a split for the
assumption of $x$, and it is not possible to split any assumption into a unique
assumption and any other assumption.

Finally, system \eqref{eq:2} above can be safely extended with processes such as
$\ptit{client}_4$ which uses unique channels obtained from the heap in
unrestricted fashion.  Our type system accepts $\ptit{client}_4$ by applying
\emph{subtyping} from unique to unrestricted on the channel $x$ obtained from
\textit{heap}.   
\begin{equation*}
\ptit{client}_4 \deftri \piin{\textit{heap}}{x}{\recX{(\pioutB{\textit{getTime}}{x}{\piinB{x}{y_\mathit{hr}, y_\mathit{min}}{P \paral X}})}}
\end{equation*}

\section{Related Work}
\label{sec:RelatedWork}

The literature on using substructural logics to support destructive or strong
updates is huge and we can give but a brief overview here. More in-depth
discussions can be found in \cite{edsko:thesis,pottier:2007}.

\begin{description}

\item[Resources and \pic]
\label{sec:resources-pic}

Resource usage in a \pic extension is studied in \cite{teller:resourcespi} but
it differs from our work in many respects.  For a start, their scoping
construct assumes an allocation semantics while we tease scoping and
allocation apart as separate constructs.  The resource reclamation construct in
\cite{teller:resourcespi} is at a higher level of abstraction than \free{c}{P},
and acts more like a ``resource finaliser'' leading to garbage collection.
Resource reclamation is implicit in \cite{teller:resourcespi}, permitting
different garbage collection policies for the same program whereas in the
\picr resource reclamation is explicit and fixed for every program.  The main
difference however concerns the aim of the type systems: our  type system
ensures
safe channel deallocation and reuse;  the type system  in
\cite{teller:resourcespi} statically determines an upper bound for the number
of resources used by a process and does not use substructural typing.

\item[Linearity versus Uniqueness]

In the absence of subtyping, affine typing and uniqueness typing coincide but
when subtyping is introduced they can be considered dual
\cite{harrington:uniquenesslogic}. For linear typing, the subtyping relation
allows coercing non-linear assumptions into a linear assumptions, \ie $!U
\rightarrow U$, but for uniqueness typing, the subtyping relation permits
coercing unique assumptions into non-unique assumptions.  Correspondingly, the
interpretation is different: linearity (\resp affinity) is a \emph{local
obligation} that a channel must be used exactly (\resp at most) once, while
uniqueness is a \emph{global guarantee} that no other processes have access to
the channel. Combining both subtyping relations as we have done in this paper
appears to be novel.  The usefulness of the subtyping relation for affine or
linear typing is well-known (e.g., see \cite{kobayashi:typesystems}); subtyping
for unique assumptions allows to ``forget'' the uniqueness guarantee;
$\ptit{client}_4$ above shows one scenario where this might be useful. 

\item[Linearity in functional programming]

In pure functional programming languages, data structures are always persistent
and destructive updates are not supported: mapping a function $f$ across a list
$[x_1, \ldots, x_n]$ yields a \textit{new} list $[f \; x_1, \ldots, f \; x_n]$,
leaving the old list intact. However, destructive updates cannot always be
avoided (\eg when modelling system I/O \cite{achten:cleanio}) and are sometimes
required for efficiency  (\eg updating arrays). Substructural type systems can
be used to support destructive update without losing referential transparency:
destructive updates are only allowed on terms that are not shared. Both
uniqueness typing \cite{barendsen:functional} and linear typing have been used
for this purpose, although even some proponents of linear typing agree that the
match is not perfect \cite[Section 3]{wadler:use}. 

In functional languages with side effects, substructural type systems have been
used to support strong (type changing) updates. For instance, Ahmed \textit{et
al.} have applied a linear type system to support ``strong'' (type changing)
updates to ML-style references \cite{ahmed:stepindexed} in a setting with no subtyping.  

It has been recognized early on that it is useful to allow the uniqueness of an
object to be temporarily violated. In functional languages, this typically
takes the form of a sequential construct that allows a unique object (such as
an array) to be regarded as non-unique to allow multiple non-destructive
accesses (such as multiple reads) after which the uniqueness is recovered
again. Wadler's \texttt{let!} construct \cite{wadler:change} (or the equivalent
Clean construct \texttt{\#!}) and observer types \cite{odersky:observers} both
fall into this category, and this approach has also been adopted by some
non-functional languages where it is sometimes called \textit{borrowing}
\cite{clarke:external}. It is however non-trivial to extend this approach to a
concurrent setting with a partial order over execution steps; our approach  can
be regarded as one preliminary attempt to do so.  

\item[Strong update in the presence of sharing]

There is substantial research on type systems that
allow strong update even in the presence of sharing; the work on alias types
and Vault \cite{smith:2000,walker:2001,fahndrich:2002} and on CQual
\cite{foster:2002} are notable examples of this. These type systems do explicit
alias analysis by reflecting memory locations at the type level through
singleton types. This makes it possible to track within the type system that a
strong (type changing) update to one variable changes the type of all its
aliases. The interpretation of unique (or linear) in these systems is
different: a unique reference (typically called a \textit{capability} in this
context) does not mean that there is only a single reference to the object, but
rather that all its aliases are known. For non-unique reference not all aliases
are known and so strong update is disallowed. 

These systems are developed for imperative languages. They are less useful for
functional languages because they cannot guarantee referential transparency,
and they appear to be even less useful for concurrent languages: even if we track the effect
of a strong update on a shared object on all its aliases, this is only useful
if we know \textit{when} the update happens. In an inherently non-deterministic
language such as the \pic this is usually hard to know before execution. 

\item[Linearity in the \pic]

Linear types for the \pic were introduced by Kobayashi \textit{et al.}
\cite{KobayashiPT:linearity} but do not employ any subtyping.  Moreover, their
system cannot be used as a basis for strong update or channel deallocation;
although they  split a bidirectional linear (``unique'') channel into
a linear input channel and a linear output channel (\cf Definition 2.3.1 for the
type combination operator $(+)$ ) these parts are never
``collected'' or ``counted''. The more refined type splitting operation we use in this paper,
combined with the type decrement operation (which has no equivalent in their
system) is key to make uniqueness useful for strong updates and deallocation.
Our system can easily be extended to incorporate modalities but it does not
rely on them; in our case, channel modalities are an orthogonal issue.

\item[Fractional permissions and permission accounting]

Boyland \cite{boyland:03fractions} was one of the first to consider splitting
permissions into \textit{fractional} permissions which allow linearity or
uniqueness to be temporarily violated. Thus, strong update is possible only
with a full permission, whereas only passive access is permitted with a
``fraction'' of a permission.  When all the fractions have been reassembled
into one whole permission, strong update is once again possible.

Boyland's suggestion has been taken up by Bornat \textit{et al.}
\cite{Bornat:05Separation}, who introduce both fractional permissions and
``counting'' permissions to separation logic. Despite of the fact that their
model of concurrency is shared-memory,  their mechanism of permission splitting
and counting is surprisingly similar to our treatment of unique assumptions.
However, while their resource reading of semaphores targets \emph{implicit}
ownership-transfer, uniqueness typing allow us to reason about explicit
ownership-transfer. Moreover, subtyping from unique to unrestricted types
provides the flexibility of not counting assumptions whenever this is not
required, simplifying reasoning for resources that are not deallocated or
strongly updated. 

\item[Session types]

Session types \cite{Honda:SessionTypes} and types with CCS-like usage
annotations \cite{kobayashi:typesystems} are used to describe channels which
 send objects of different types. However, these types give 
detailed information on how channels are used, which makes modular typing
difficult. For example, the \textit{heap} channel used by $\ptit{client}_3$
cannot be given a type without knowing all the processes that use the heap. 

\end{description}

\section{Conclusions and Future Work}

We have extended ideas from process calculi, substructural logics and
permission counting to define a type system for a the pi-calculus extended with
primitives for channel allocation and deallocation, where strong update and
channel deallocation is deemed safe for unique channels. 

The purpose of our type system is not to ensure that every resource that is
allocated will also be deallocated (\ie the absence of memory leaks). This is difficult to track in a type
system. For instance, consider
\begin{equation*}
\alloc{x}{\bigl(\;\piout{c}{d_1}{\pioutB{d_1}{}{\inert}} \quad\paral\quad \piout{c}{d_2}{\inert} \quad \paral\quad \piin{c}{y}{\piinB{y}{}{\freeA{x}}}\;\bigr)} 
\end{equation*}
Statically, it is hard to determine whether the third parallel process will
eventually execute the $\freeA{x}$ operation.  This is due to the fact that it
can non-deterministically react with either the first or second parallel
process and, should it react with the second process, it will block at
\piinB{d_2}{}{\freeA{x}}.  In order to reject this process as ill-typed, the
type-system needs to detect potential deadlocks.  This can be done
\cite{kobayashi:2006}, but requires a type system that is considerably more
complicated than ours. We leave the responsibility to deallocate to the user,
but guarantee that resources once deallocated will no longer be used. 

The simplicity of our type-system makes it easily extensible.  For instance,
one useful extension would be that of input/output modalities, which blend
easily with the affine/unique duality.  Presently,   when a server process
splits a channel $c : \chantyp{\tV}{\uniqueNow}$ into one channel of type
$\chantyp{\tV}{\unique{2}}$ and two channels of type $\chantyp{\tV}{\affine}$
to be given to two clients, the clients can potentially use this channel to
communicate amongst themselves instead of the server.  Modalities are a natural
mechanism  to preclude this from happening.

We are currently investigating ways how uniqueness types can be used to refine
existing equational theories, so as to be able to equate processes such as
$\ptit{client}_1$ and $\ptit{client}_0$.   This will probably require us to
establish a correspondence between uniqueness at the type level and restriction
at the term level.

\end{document}